\documentclass[floatfix,aps,pre,twocolumn]{revtex4}  
\usepackage{graphicx,amsmath}
\usepackage[amssymb]{SIunits}
\usepackage{color,xcolor}
\usepackage{placeins}

\definecolor{lightred}{RGB}{255,255,255}
\newcommand{\makered}[1]{\textcolor{black}{#1}}

\begin{document}

\title{The Boiling Twente Taylor-Couette (BTTC) facility: temperature controlled turbulent flow between independently rotating, coaxial cylinders}
\author{Sander G. Huisman}  
\email{s.g.huisman@gmail.com}
\author{Roeland C. A. van der Veen}  
\email{r.c.a.vanderveen@utwente.nl}
\author{Gert-Wim H. Bruggert}
\author{Detlef Lohse}
\email{d.lohse@utwente.nl}
\author{Chao Sun}
\email{c.sun@utwente.nl}
\affiliation{Department of Applied Physics and J. M. Burgers Centre for Fluid Dynamics, University of Twente, P.O. Box 217, 7500 AE Enschede, The Netherlands}

\date{\today}

\newcommand{\tc}{Taylor-Couette}
\newcommand{\tttc}{T${}^3$C}

\begin{abstract} 
A new \tc~system has been designed and constructed with precise temperature control. Two concentric independently rotating cylinders are able to rotate at \makered{maximum} rates of $f_i = \pm\unit{20}{\hertz}$ for the inner cylinder and $f_o = \pm\unit{10}{\hertz}$ for the outer cylinder. The inner cylinder has an \makered{outside} radius of $r_i = \unit{75}{\milli \meter}$, and the outer cylinder has an \makered{inside} radius of $r_o = \unit{105}{\milli \meter}$, resulting in a gap of $d=\unit{30}{\milli \meter}$. The height of the gap $L = \unit{549}{\milli \meter}$, giving a volume of $V=\unit{9.3}{\liter}$. The geometric parameters are $\eta = r_i/r_o = 0.714$ and $\Gamma = L/d = 18.3$. With water as working fluid at room temperature the Reynolds number\makered{s} that can be achieved are \makered{$\text{Re}_i = \omega_i r_i (r_o-r_i)/\nu = 2.8 \times 10^5$} and \makered{$\text{Re}_o = \omega_o r_o (r_o-r_i)/\nu = 2 \times 10^5$}, or a combined Reynolds number of \makered{up to $\text{Re}  = (\omega_i r_i -\omega_o r_o)(r_o-r_i)/\nu = 4.8 \times 10^5$}. If the working fluid is changed to the fluorinated liquid FC-3284 with kinematic viscosity \unit{0.42}{\centi St}, the combined Reynolds number can reach $\text{Re} = 1.1 \times 10^6$. The apparatus features precise temperature control of the outer and inner cylinder separately, and is fully optically accessible from the side and top. The new facility offers the possibility to accurately study the process of boiling inside a turbulent flow and its effect on the flow. 
\end{abstract}

\maketitle

\section{Introduction}

Though well known in the context of rheology, the \tc~geometry---a geometry in which fluid is bound between two concentric rotating cylinders---has also been used in many fundamental concepts: the verification of the no-slip boundary condition, hydrodynamic stability \cite{tay23}, higher and lower order bifurcation phenomena and flow structures \cite{fenstermacher1979, diprima1985, and86, huisman2014a}, but also in the field of combustion \cite{aldredge1996, aldredge1998, ronney1995}, drag reduction \cite{berg2005a, sug08a, gils2013a,mckinley2015}, magnetohydrodynamics in order to study \textit{e.g.}~the MRI \cite{chandrasekhar, balbus1991, rudiger2001, ji2001, hollerbach2010}, astrophysics to study Keplerian flow in accretion discs \cite{richardzahn1999,dub05,schartman2012,pao12a}, rotating filtration in order to extract plasma from whole blood\cite{ohashi1988,beaudoin1989,ameer1999,wereley1999,serre2008}, cooling of rotating machinery \cite{jeng2007}, flows in bearings, the fundamentals of high Reynolds number flows \cite{lewis1999, gils2011a, paoletti2011, huisman2013, huisman2013a, ostilla2014, ostilla2014a, ostilla2014b, huisman2014a}, and as a catalytic and plasmapheretic reactor \cite{cohen1991, ameer1999b, ameer1999ex}. Hence the \tc~geometry is a versatile geometry in physics, engineering, and beyond. 

Despite its ubiquity in industrial applications and our daily life, boiling and cavitation, collectively constituting rapid liquid-to-vapor phase transitions (as opposed to slow evaporation), is a field hitherto unexplored in \tc. Withal, boiling has almost exclusively been an engineering subject. \makered{However, the nucleation of the bubble, the interaction of the gas and vapour with the liquid (absorption, evaporization, and heat transfer), the fluid dynamics of the vapour and gas inside the nucleating bubble and the fluid dynamics of the liquid around the oscillating, moving, and growing bubbles are hardly touched in the field of physics and only ``quasi-static boiling'' is considered, \textit{i.e.~}the statistics and thermodynamics of phase transitions. Nevertheless, above mentioned rapid features of the phase transitions---boiling---are certainly interesting and involve practically relevant physics} that is still in need of deep and fundamental understanding. Boiling can be studied at the surface and in the bulk \cite{lakkaraju2011,lakkaraju2013a}, and can focus on a single vapor (nano)bubble \cite{limbeek2013a,zhang2014a}, or the collective effects of multiple vapor bubbles. The complexity and intractability of the boiling process is very well exemplified by the so-called boiling curve, which shape and explanation is far from trivial \cite{dhir1998, theofanous2002boiling, dhir2006, kim2009}, and still lacks detailed physical interpretation.

With the TC system introduced in this paper we plan to study the collective dynamics of many vapor bubbles in the bulk \cite{zhong2009}. The combination of vapor bubbles with turbulence complicates the phase transition: first, the present vapor bubbles provide nucleation sites for more boiling, second, the turbulence provides intermittent flow (with associated intermittent kinetic energy spikes), and third, turbulent flows have locally low pressure spots---all of which provide favorable conditions for boiling to initiate.

To trigger the liquid-to-gas phase transition we have to carefully control the \makered{kinetic and thermal} energy of our working liquid. It is therefore necessary to have a closed geometry with precise boundary conditions for both the velocity (kinetic energy) and the temperature (thermal energy). The \tc~geometry is a good candidate to study the physics of boiling in a controlled way because we can control these boundary conditions, it has global \makered{energy and momentum} balances\makered{\cite{eckhardt2007}}, its geometry has many symmetries, and it is relatively easy to construct. Though our existing apparatus, the \tttc\cite{gil11b}, is able to accurately rotate and control the temperature within \unit{0.1}{\kelvin}, its rather large volume of \unit{111}{\liter} makes an experiment with low-boiling-point fluorinated liquids like FC-87 or FC-3284 a costly endeavour. The  presented apparatus is smaller compared to the \tttc~facility, but is still \makered{capable of attaining a very turbulent flow of $\text{Re}=\omega_i r_i (r_o-r_i)/\nu= \mathcal{O}(10^6)$} and operates in the so-called ultimate regime \cite{eckhardt2007} so the study of nucleation of vapor bubbles inside Taylor vortices can be performed. For more details about the ultimate regime, and an introduction in Taylor-Couette flows in the ultimate regime in context of the applicability of the \tc~geometry we refer the reader to the review article of Grossmann, Lohse, and Sun\cite{lohse2016arfm}. \makered{The classical (non-ultimate) regime, with its many different flow structures, is well described by the pioneering work of Andereck \text{et al.~}\cite{and86} and by a recent review article by Fardin \textit{et al.~}\cite{fardin2014}.} Other setups \makered{for high Re flows} have been created \cite{lathrop1992, schartman2009, merbold2013, kerstin2013}, but either do not feature the very precise rotation, and cooling and heating methods that we aim for, or do not have completely transparent top and side cylinders, or do not feature an overflow channel that we need for vaporizing liquids. 

The new setup is designed to study macroscopic boiling in the bulk of the working fluid. \makered{The improved design is based on the} \tttc \makered{\cite{gil11b}}, and includes accurate temperature control, and is dubbed the Boiling Twente Taylor-Couette (BTTC) facility. \makered{The system is able to accurately heat and cool both the inner and outer cylinder independently over a large temperature range of \unit{0}{\celsius}--\unit{60}{\celsius} and can be extended using special coolants rather than water. The temperature of the working fluid is closely monitored at three locations on the inner cylinder and at two locations at the outer cylinder. Section \ref{lab:temperature} describes the temperature control in more detail. The inner and outer cylinder have a maximum rotation rate of $f_i = \pm \unit{20}{\hertz}$ and $f_o = \pm \unit{10}{\hertz}$, respectively. The maximum Reynolds numbers are tabulated in \ref{table:maxnumbers}, and more details about the driving mechanism can be found in section \ref{sec:driving}. The inner cylinder comprises three, vertically stacked, cylinders. The top and bottom inner cylinders can be removed in order to have vertical gaps of similar width \makered{as} the radial gap. The entire outer cylinder and the top plate rotate together and are completely transparent, and do not have obstructing tension bars like the \tttc \cite{gil11b}. These transparent parts allow for detailed investigation of local quantities using optical methods like PIV, LDA, and PTV, and regular (high-speed) imaging. An overflow channel that can be used during rotation of either cylinder is one of the unique features of the presented setup, and will be used to capture the expanding or boiling fluid and vapor.}

\begin{table}[hptb!]
  \centering
    \caption{\makered{Current maximum rotation rates and their corresponding Reynolds number of the BTTC around room temperature.}}
    \colorbox{lightred}{
 		\begin{tabular}{|r|c|c|c|}
 		\hline
& Water & FC-3248 & FC-87\\ [1mm]
\hline
$f_i$ & \multicolumn{3}{|c|}{$\pm \unit{20}{\hertz}$}  \\
\hline
$f_o$ & \multicolumn{3}{|c|}{$\pm \unit{10}{\hertz}$} \\
\hline
$\text{Re}_i = \omega_i r_i (r_o - r_i ) /\nu$ & $2.8 \times 10^5$ & $6.7 \times 10^5$ & $10.5 \times 10^5$ \\
\hline
$\text{Re}_o = \omega_o r_o (r_o - r_i ) /\nu$ & $2.0 \times 10^5$ & $4.7 \times 10^5$ & $7.3 \times 10^5$ \\
\hline
$\text{Re}  = \text{Re}_i + \text{Re}_o$ & $4.8 \times 10^5$ & $11.4 \times 10^5$ & $17.8 \times 10^5$ \\
\hline
		\end{tabular}
		}
  \label{table:maxnumbers}
\end{table}

\section{System description}
\subsection{General}

\begin{figure}[hpbt!]
	\begin{center}
		\includegraphics[width=0.8 \linewidth]{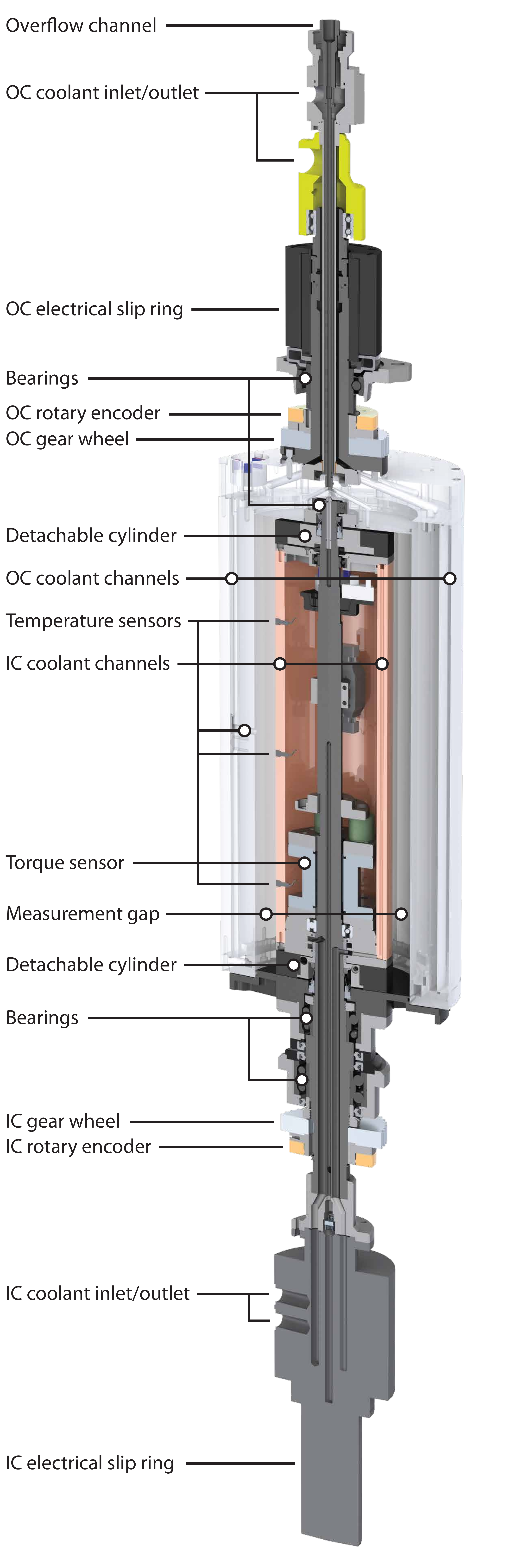}
		\caption{Vertical cross section of the BTTC. The inner cylinder is made of copper and is designated as \textsf{IC}. The outer cylinder, designated as \textsf{OC}, is made of PMMA. All the steel that is in contact with water is made of 316 grade stainless steel to prevent corrosion, except for the axis of the inner cylinder which is machined from Duplex 1.4462 stainless steel.}
		\label{fig:cross}
	\end{center} 
\end{figure}

\begin{table}[hptb!]
  \centering
    \caption{Geometric dimensions of the BTTC and the \tttc~facility \cite{gil11b}. The outer cylinder of the \tttc~can be varied and all flow options are listed.}
 		\begin{tabular}{|c|c|c|c|c|c|}
 		\hline
& BTTC & \multicolumn{4}{|c|}{\tttc}  \\ [1mm]
\hline
$L$ (\unit{\!\!}{\milli \meter}) & 549 & \multicolumn{4}{|c|}{927} \\
$L_{\text{middle}}$ \makered{(\unit{\!\!}{\milli \meter})} & 489 & \multicolumn{4}{|c|}{536} \\
$L_{\text{top/bottom}}$ \makered{(\unit{\!\!}{\milli \meter})} & 26 & \multicolumn{4}{|c|}{193.5} \\
$r_i$ (\unit{\!\!}{\milli \meter}) & 75 & \multicolumn{4}{|c|}{200} \\
\cline{3-6}
$r_o$ (\unit{\!\!}{\milli \meter}) & 105 & 279 & 260 & 240 & 220 \\
$d = r_o - r_i$  (\unit{\!\!}{\milli \meter}) & 30 & 79 & 60 & 40 & 20 \\
$\eta = r_i/r_o$ & 0.714 & 0.716 & 0.769 & 0.833 & 0.909 \\
$\Gamma = L/d$ & 18.3 & 11.68 & 15.45 & 23.18 & 46.35 \\
$V$ (\unit{\!\!}{\liter}) & 9.31 & 111 & 80.4 & 51.3 & 24.5 \\
\hline
		\end{tabular}
  \label{table:dimensions}
\end{table}

The system has two independently rotating cylinders confining a liquid in the measurement gap, see figure \ref{fig:cross} for a cross section of the BTTC. The inner cylinder is made of copper and is chrome plated and has an \makered{outside} radius $r_i = \unit{75}{\milli \meter}$. The outer cylinder is made from Poly(methyl methacrylate) (PMMA) and has an \makered{inside} radius of $r_o = \unit{105}{\milli \meter}$. The radial gap is therefore $d = r_o - r_i = \unit{30}{\milli \meter}$, and the radius ratio is $\eta = r_i/r_o = 0.714$. The gap has a height of $L = \unit{549}{\milli \meter}$, giving a measurement volume of $V = \pi (r_o^2-r_i^2) L = \unit{9.31}{\liter}$, and an aspect ratio of $\Gamma = L /d = 18.3$. The middle inner cylinder, see fig.~\ref{fig:cross}, has a height of $L_{\text{middle}} = \unit{489}{\milli \meter}$, while the height of the top and bottom inner cylinders are $L_{\text{top/bottom}} = \unit{26}{\milli \meter}$. The gap between each of the cylinders and between the top and the bottom are $\delta = \unit{2}{\milli \meter}$. In addition, top and bottom inner cylinders have been made that leave no gap between the inner cylinders. Table \ref{table:dimensions} summarizes the relevant dimensions. The top plate and the outer cylinder are machined from transparent PMMA by Hemabo (Hengelo, The Netherlands) and are heat treated to relieve internal stresses in the material, while the bottom plate is machined from corrosion-resistant 316 stainless steel. These end plates and the outer cylinder are fixed together and rotate in unison. The driving axis of the inner \makered{cylinder is machined} from Duplex 1.4462 stainless steel, while the end plates of the middle inner cylinder, as well as the holders for the detachable cylinders are made from 316 stainless steel. \makered{The driving as well as the ingress and egress of the coolant of the inner cylinder is at the bottom, while these are situated at the top for the outer cylinder, see also fig.~\ref{fig:cross}.} The bearing in the top-plate is sealed using a mechanical seal (AES Seal, type: \texttt{B052-20}), and the measurements gap is sealed at the bottom by another mechanical seal (AES Seal, type: \texttt{B052-40}), sealing the interface between the outer and inner cylinder. Both mechanical seals are of the carbon-ceramic type. At the top of the middle inner cylinder a V-ring seal (type: \texttt{V-25S}) is employed. The top and bottom inner cylinders are not hollow and therefore do not require sealing. The bearings of  the driving axis of the inner cylinder, and the bearings of the bottom plate of the outer cylinder are protected against dust and water using seals (Simmering, \texttt{BAUX2 NBR}).

\subsection{Driving} \label{sec:driving}
\begin{figure}[hpbt!]
	\begin{center}
		\includegraphics[width=0.99 \linewidth]{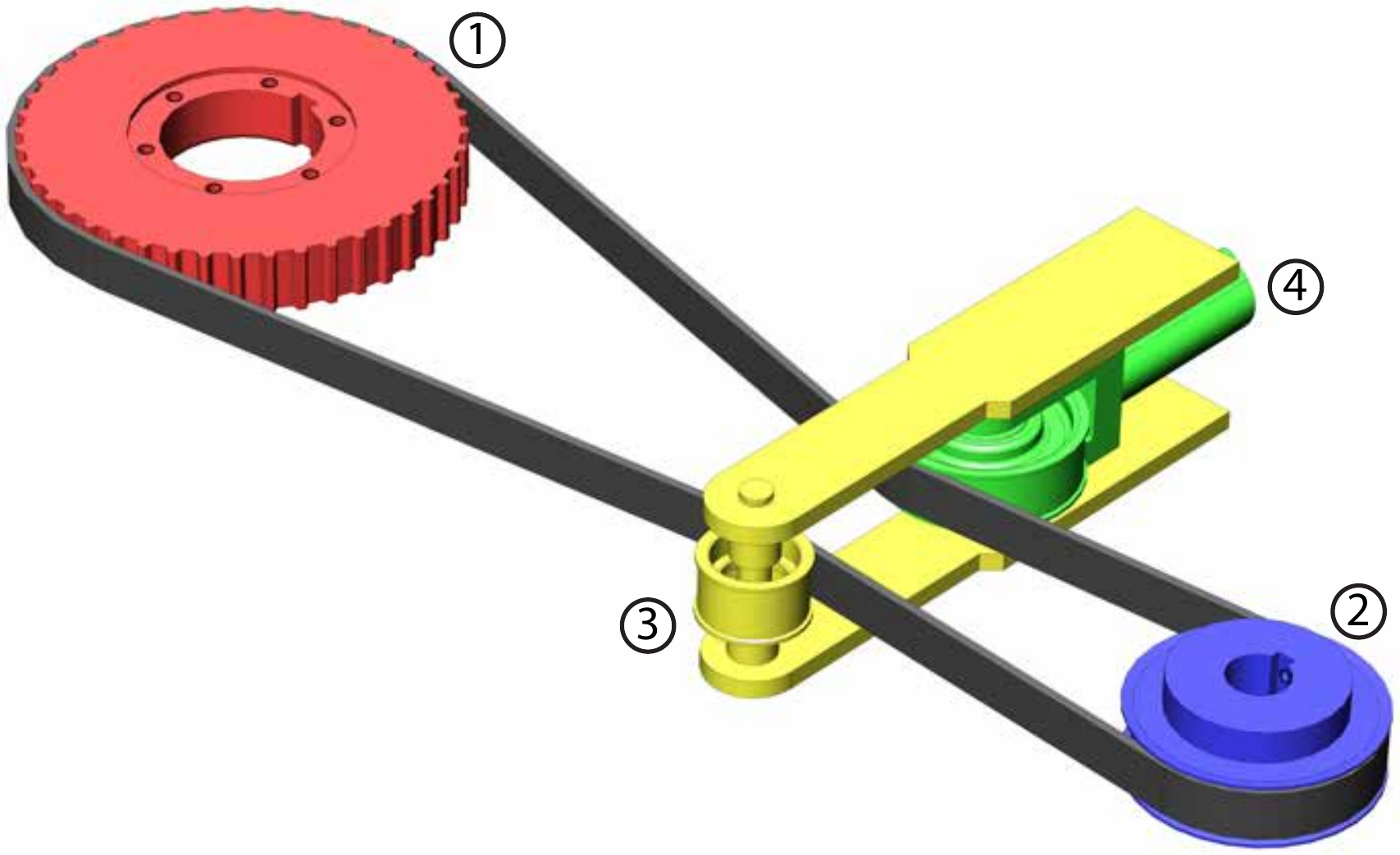}
		\caption{Inner (lower) pulley assembly. The red pulley \makered{(1)} is attached to the axis driving the inner cylinder, the blue pulley \makered{(2)} is attached to driving motor. In yellow \makered{(3)} is the idler pulley, while the belt tensioner (green\makered{, 4}) can be moved inwards/outwards to adjust the tension. Keyed joints and toothed pulleys do not allow for differential rotation between the belt, pulley, and their respective axes.}
		\label{fig:driveassembly}
	\end{center} 
\end{figure}

The inner (outer) cylinder was designed to rotate at the current maximum rotational velocity of \unit{20}{\hertz} (\unit{10}{\hertz}), although there is room to increase those velocities if the vibrations in the system, forces on the outer cylinder, and heating up of the slip rings allow for that. The inner cylinder has a gear ratio of $44\!:\!22=2\!:\!1$ while the outer cylinder has a ratio of $44\!:\!18\approx 2.44\!:\!1$. Both the inner and outer cylinder are driven by \unit{16}{\milli \meter} toothed belts and pulleys from M\"adler, Germany (type: \texttt{AT10}), and each belt has an idler pulley and a belt tensioner to ensure the belt neither vibrates nor skips any teeth. Figure~\ref{fig:driveassembly} shows the pulley assembly of the inner cylinder, the assembly of the outer cylinder is nearly identical. All pulleys feature keyways and all driving axes feature keyseats preventing relative rotation. Both cylinders are driven by servomotors with internal feedback resolvers from Beckhoff (type \texttt{AM8543-1H00} for the IC, type \texttt{AM8543-1E00} for the OC) and have a rated power of \unit{2.57}{\kilo \watt} and \unit{1.39}{\kilo \watt} and maximum rotation rate of \unit{5000}{rpm} and \unit{2500}{rpm} for the inner and outer cylinder engine, respectively. 

Both servomotors are driven by a 2-channel digital compact servo drive from Beckhoff (type \texttt{AX5206}), and each of the outputs of this drive is filtered by a \texttt{AX2090-MD50-0012} motor choke from Beckhoff. The drive is outfitted with a brake resistor (Beckhoff, up to \unit{600}{\watt} dissipation) to speed up the braking. The drive is controlled through high-speed EtherCAT system communication, and is connected to the other programmable logic controllers (PLCs). More about system control can be found in section \ref{sec:wiring}.

Magnetic angle encoders are used to independently measure the angle of the shafts, and, by differentiation, the rotation rates. Each shaft is outfitted with an \texttt{ERM200}-series angle encoder from Heidenhain (Schaumburg, USA) with a bore of \unit{70}{\milli \meter}, and a line count of 900, giving an angular resolution of \unit{0.4}{\degree} with an accuracy of \unit{\pm 25}{\arcsecond}. These magnetic encoders are connected to two incremental encoder interfaces by Beckhoff (type \texttt{EL5101}).

\begin{figure}[hbpt!]
	\begin{center}
	\colorbox{lightred}{
		\includegraphics{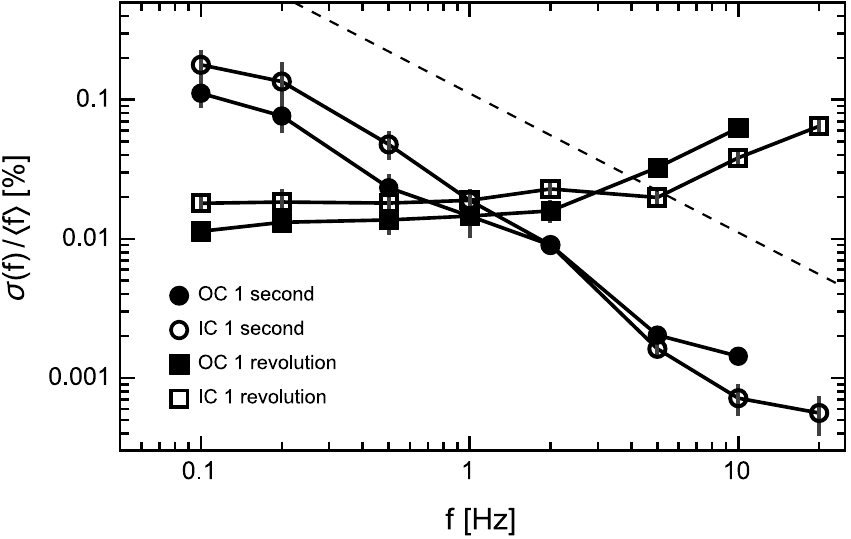}
		}
		\caption{Standard deviation of the inner cylinder (\textsf{IC}) and outer cylinder (\textsf{OC}) rotation rates expressed as a percentage of the average rotation rate. We extract the \makered{mean} velocity over a period of 1 second or the equivalent time of 1 revolution. \makered{The dashed line shows the maximum error that can originate from the angular resolution of the magnetic encoders for the case in which we average over 1 second. The slope of this line is $-1$.}}
		
		\label{fig:rotation}
	\end{center} 
\end{figure}

The performance of our driving system is checked by varying the rotation rate from \unit{0.1}{\hertz} to \unit{20}{\hertz} \makered{for the inner cylinder} and from \unit{0.1}{\hertz} to \unit{10}{\hertz} \makered{for the outer cylinder.} The angle was measured at an acquisition rate of \unit{100}{\hertz}, and using a sliding linear fit the rotation rate is extracted for a duration of either 1 second or a period corresponding to 1 revolution. The standard deviation is then computed for these rotation rates, and normalized by the average rotation rates, see fig.~\ref{fig:rotation}. The rotation rate always has a standard deviation smaller than $0.2\%$ of the setpoint, and generally much better than this figure. \makered{Note that for measurements where the rotation angle is small (slow rotation speeds during a 1 second interval), the angular resolution of \unit{0.4}{\degree} can cause errors. This maximum error caused by the angular encoder is indicated in fig.~\ref{fig:rotation}, and is based on a measurement duration of 1 second. To exemplify: if we rotate at a frequency of \unit{\frac 59}{\hertz} and we do a measurement over 1 second, the cylinder should have rotated \unit{200}{\degree}. The angular resolution of \unit{0.4}{\degree} means that both the initial and final positions are both at most \unit{0.2}{\degree} off from their actual values, and the error caused by the finite resolution is therefore at most (worst-case scenario) $(\unit{0.2}{\degree}+\unit{0.2}{\degree})/\unit{200}{\degree} = 0.2\%$. We find that the deviations for the 1 second interval measurements are below this limit indicating that the error is most likely caused by the finite resolution of the measurement in addition to the error introduced by the engine.} We compare the performance from the current setup with the \tttc\cite{gil11b} and the new \tc~setup in G\"ottingen\cite{kerstin2013} for which the authors compare the standard deviation for the case of $f=\unit{5}{\hertz}$ over a single cylinder revolution. They found that they were both $0.1\%$. Our current setup has values of $0.02\%$ and $0.03\%$ (at \unit{5}{\hertz}) for the inner and outer cylinders, respectively. These standard deviations increase to about $0.06\%$ for the maximum rotation rates of both cylinders.

\subsection{Temperature} \label{lab:temperature}

\begin{figure}[hbtp!]
	\begin{center}
		\includegraphics{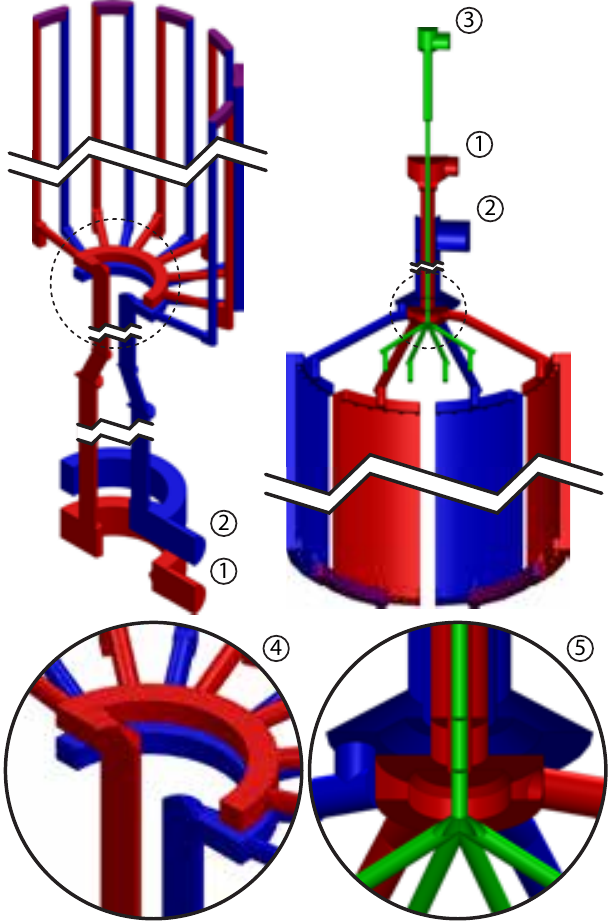}
		\caption{Left(right): Section view of the cooling channels of the inner(outer) cylinder. Coolant enters the cooling circuit at \textsf{1} where the fluid goes through a rotary union to the rotating frame. The coolant is split up in to 12(4) parallel streams inside the inner(outer) cylinder, see the detail view \textsf{4}(\textsf{5}) at the bottom. The coolant then flows axially, enters the purple `bridges' and returns in the other 12(4) parallels channels, and is again combined at \textsf{4}(\textsf{5}). The returning coolant continues to flow axially (and concentrically for the outer cooling circuit) to the entering coolant. The coolant leaves the coolant circuit through a rotary union at \textsf{2}. The apparatus is equiped with an overflow channel, shown in green. A mix of vapor, gas, and liquid can exit the system through the rotary union at \textsf{3}.}
		\label{fig:cooling}
	\end{center} 
\end{figure}

\begin{table}[htpb!]
  \centering
    \caption{Standard deviation of the local temperatures for a low setpoint around \unit{20}{\celsius} and a high setpoint around \unit{50}{\celsius}. The duration of each of the recordings is one hour long. $\sigma(\hdots)$ denotes standard deviation, $\langle \hdots \rangle$ denotes averaging, and the subscript $n$($t$) denotes that the operation is over the different sensors(over time). The time evolution for the entry with a star is shown in fig.~\ref{fig:temperature}.}
 		\begin{tabular}{|l|r|c|c|}
 		\hline
 Case & $f_i$ [Hz] & $\left\langle \sigma _t\text{(T)} \right \rangle_n$ [K] & $\sigma_n(\left \langle T \right \rangle_t)$ [K] \\
  \hline
 Low & 1 & 0.013 & 0.056 \\
 Low & 2 & 0.012 & 0.055 \\
 Low* & 5 & 0.013 & 0.052 \\
 Low & 10 & 0.015 & 0.049 \\
 Low & 20 & 0.014 & 0.054 \\
 High & 1 & 0.017 & 0.088 \\
 High & 2 & 0.019 & 0.042 \\
 High & 5 & 0.027 & 0.051 \\
 High & 10 & 0.021 & 0.062 \\
 High & 20 & 0.021 & 0.061 \\
 \hline
		\end{tabular}
  \label{table:temperature}
\end{table}

\begin{figure}[hbpt!]
	\begin{center}
		\includegraphics[width=0.99 \linewidth]{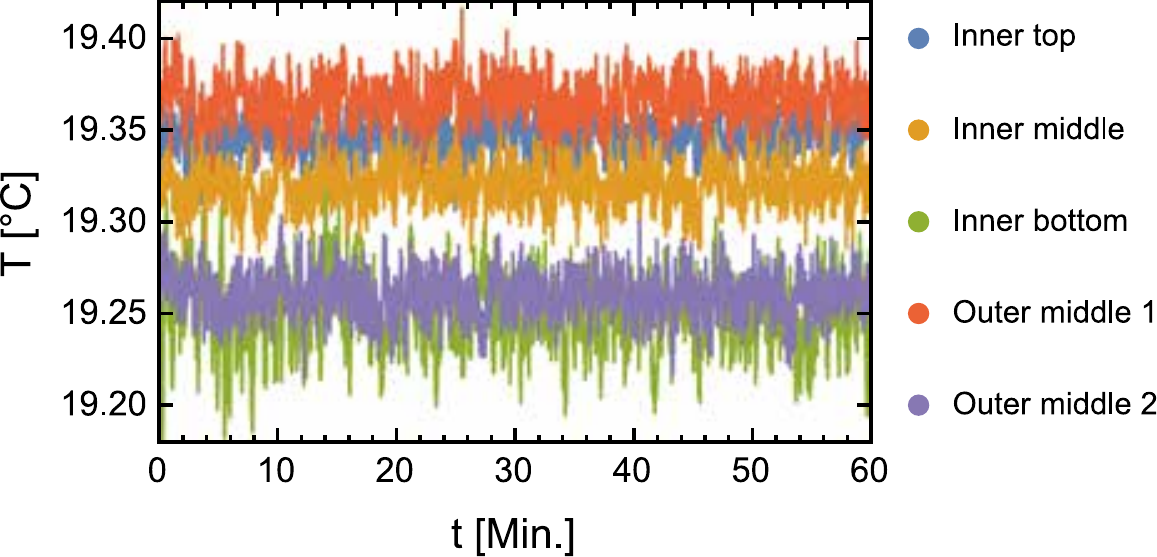}
		\caption{Time evolution of the temperature for the low temperature setpoint for the case of $f_i = \unit{5}{\hertz}$, see the starred row in table \ref{table:temperature}. The colors correspond to different sensors in either the inner or outer cylinder, see the legend on the right.}
		\label{fig:temperature}
	\end{center} 
\end{figure}

\begin{figure*}[hbtp!]
	\begin{center}
		\includegraphics[width=0.75 \linewidth]{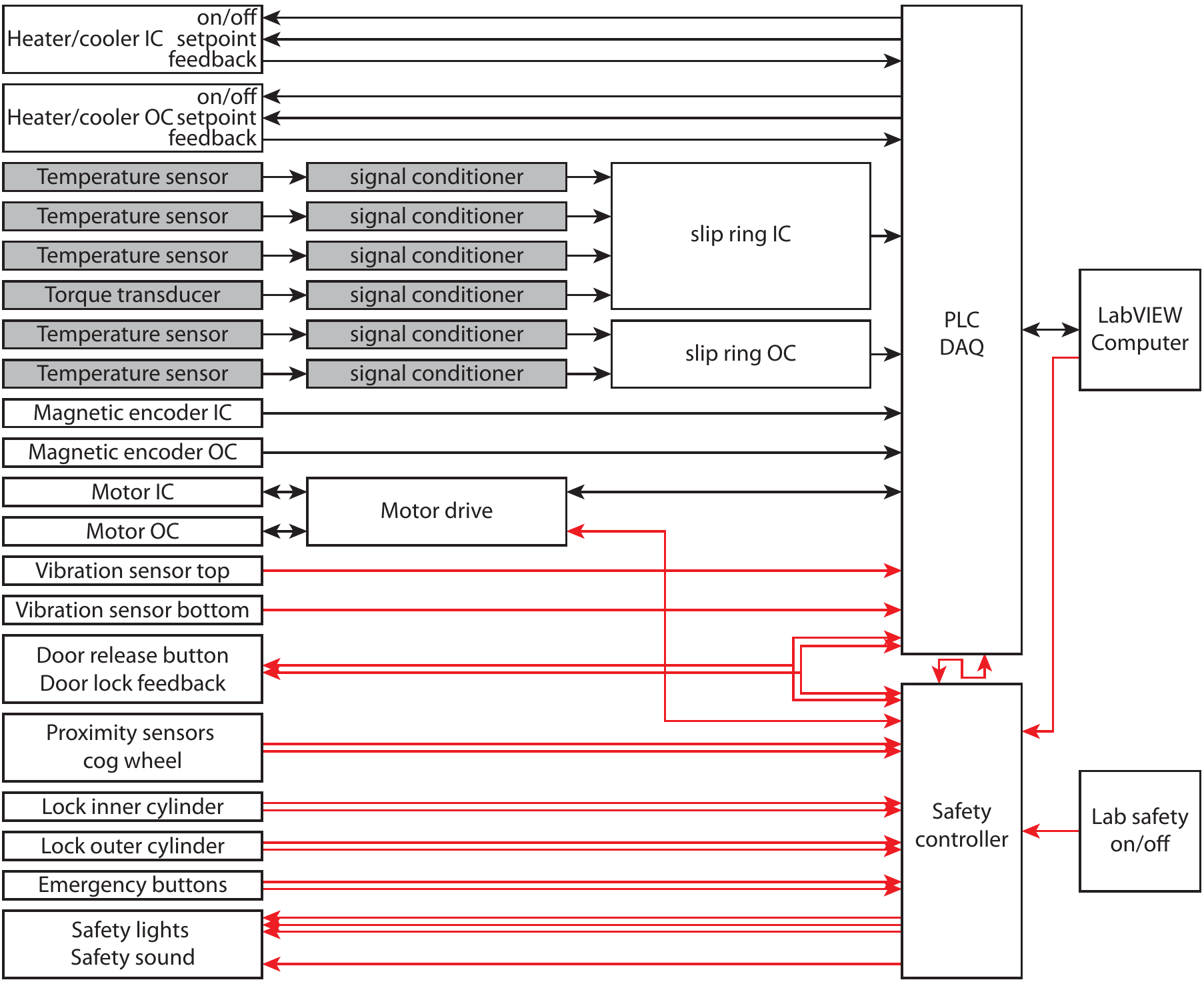}
		\caption{Sketch of the BTTC wiring. Grey components are embedded in the inner or outer cylinder. Red lines indicate that these control safety-related aspects of the setup. PLC stands for programmable logic controller, DAQ for data acquisition, IC for inner cylinder, and OC for outer cylinder.}
		\label{fig:circuitry}
	\end{center} 
\end{figure*}

To allow for experiments with a very precise control of the temperature we have chosen to cool/heat the working fluid from the inner cylinder as well as from the outer cylinder in separate cooling circuits. In order to maximize the cooling power, the inner cylinder has been machined of copper, and has been chrome plated in order to prevent corrosion. The inner cylinder has 12 parallel channels (and 12 return channels) for the coolant, making use of countercurrent heat exchange to lower thermal inhomogeneities, see also fig.~\ref{fig:cooling}. The outer cylinder is machined from PMMA (Hemabo, Hengelo, The Netherlands) and it therefore has a poor thermal conductivity, despite that, we also cool the outer cylinder in a second cooling circuit that goes through the driving axis of the outer cylinder (at the top) in order to ensure that the thermal boundary is always identical. The outer cylinder has 4 parallel streams where the coolant flows axially around the working fluid inside the measurement gap. Four rotary unions provide the means of ingress to and egress from the cooling circuits (bottom: GAT mbH, Germany, type \texttt{Rotocombi SW65-W2-HM}, top: Deublin, \texttt{357-130-235} and a custom-made rotary union). An additional rotary union (Deublin, type: \texttt{1115-000-205}) is installed at the top, and allows for the system to overflow such that the expanding liquid, gas, or vapor can be captured, see the green channels in fig.~\ref{fig:cooling}. The cooling circuit of the inner and outer cylinder are connected to two refridgerated/heating circulators by Julabo (type: \texttt{FP50-HL}) with a maximum cooling capacity \unit{0.9}{\kilo \watt}, a heating capacity of \unit{2}{\kilo \watt}, a temperature stability of \unit{0.01}{\kelvin}, and a working temperature range of \unit{-50}{\celsius} -- \unit{200}{\celsius}. Our temperature range is however limited by the \makered{glass transition temperature} of PMMA \makered{which can be as low as \unit{82}{\celsius}}, and the heaters are therefore limited to a maximum of \unit{60}{\celsius} to ensure full strength of the outer cylinder during rapid rotation \makered{while having a large safety margin}. The Julabo coolers/heaters now have water as their coolant, in order to reach sub-zero temperatures for other types of experiments one has to replace the coolant by \textit{e.g.}~a glycol based or glycerol based coolant.

The inner cylinder embeds three \texttt{PT100} temperature sensor at different heights to monitor the temperature of the working fluid, while the outer cylinder embeds another two sensors to check the temperature. The five \texttt{PT100} sensors are each connected to a precalibrated signal conditioner IPAQ-H$^{\text{plus}}$ from Inor (Malm\"o, Sweden) inside the inner or outer cylinder. This combination of sensor and signal conditioner results in a specified absolute temperature accuracy of \unit{0.1}{\kelvin} and for a relative temperature accuracy of \unit{0.01}{\kelvin}. The three sensors inside the inner cylinder are mounted flush (at $z=0.22L$, $z = 0.49L$, and $z = 0.76L$), are thermally isolated from the copper, and are in direct contact with the working fluid. The two sensors in the outer cylinder are also mounted flush and are both located at $z=0.53L$ and are positioned \unit{180}{\degree} apart for balancing reasons. The current signals from the signal conditioners exit the system through slip rings, see fig.~\ref{fig:cross}, and are read out by 4-channel analog input terminals from Beckhoff (type \texttt{EL3154}).

The performance of our temperature control is checked by varying the rotation rate and the setpoint: a low temperature around room temperature and a high temperature around \unit{50}{\celsius}. We tabulate the stability of the temperature in tab.~\ref{table:temperature}. We find that the temperature stability is better than \unit{0.03}{\kelvin}, and generally better than \unit{0.02}{\kelvin}, see the 3${}^{\text{rd}}$ column of tab.~\ref{table:temperature}. The temperature distribution inside the measurement gap (4${}^{\text{th}}$ column) is better than \unit{0.1}{\kelvin}, which is in correspondence with the absolute temperature accuracy of the sensors of \unit{0.1}{\kelvin}. As an example we show the time evolution of the temperature for the low temperature setpoint for the case of $f_i = \unit{5}{\hertz}$, see fig.~\ref{fig:temperature}. We do not find any systematic radial or height-dependence of the temperature due to the large cooling/heating surface (the entire inner and outer cylinder) combined with the excellent turbulent mixing causing a nearly homogeneous temperature inside the measurement gap, which provides an excellent playground to study the physics of boiling in the bulk.

\subsection{Wiring and control} \label{sec:wiring}
A diagram of the BTTC wiring is shown in fig.~\ref{fig:circuitry}. Temperature and torque signals from the inner cylinder leave the rotating frame through a slip ring at the bottom made by GAT mbH (type: \texttt{SRK 80/18}, 18 channels). The temperature signals from the outer cylinder exit the system at the top via a through-bore slip ring (type: \texttt{SRH 80180 FT 14}, 14 channels) made by Gileon. The signals are digitalized by analog input modules (type \texttt{EL3154} and \texttt{EL3124}) from Beckhoff. The communication to the Julabo temperature controllers are done through analog output modules (type: \texttt{EL4004}) from Beckhoff. The communication with all the DAQ and PLCs is done through the EtherCAT protocol (by Beckhoff Automation), which connects the computer and the Beckhoff modules and the communication to the safety module from Omron (type \texttt{G9SX-LM244-F10}). See section \ref{sec:safety} for more information about safety. All the data acquisition and control of the coolers/heaters and motors are performed at a rate of \unit{100}{\hertz} by the computer. The computer runs a LabVIEW (National Instruments) program by which the end user can control the various peripherals and save all the data. The LabVIEW software allows the end user to \textit{e.g.~}make a continuous velocity ramp from an initial rotational velocity to a final rotational velocity over a certain amount of time, for both cylinders individually. Also tabulated entries can be imported. Each entry will be performed sequentially and can contain different rotational velocities, cooling temperatures, filename for saving et cetera. These automated features minimize operational errors while maintaining high safety standards, see section \ref{sec:safety} for more about safety. 

\begin{figure}[hbpt!]
	\begin{center}
		\includegraphics[width=0.99 \linewidth]{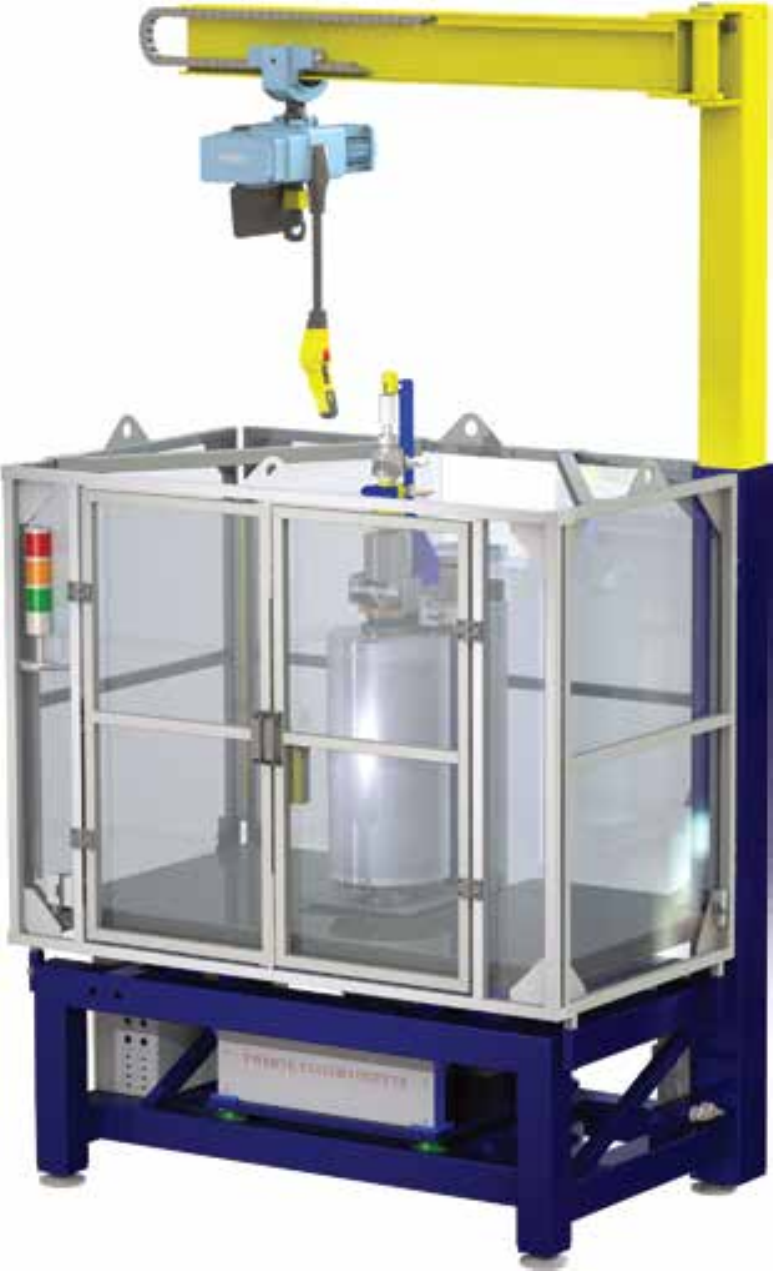}
		\caption{Rendering of the new BTTC apparatus.}
		\label{fig:rendering}
	\end{center} 
\end{figure}

\subsection{Safety features} \label{sec:safety}

The systems employs a wide variety of safety features to allow for a safe operation of the apparatus. The LabVIEW software continuously monitors the vibrations using vibration sensors by Sensonics Ltd (type: \texttt{PZDC-58K00110}, 0--\unit{15}{\milli \meter \per \second} range), mounted at the top and bottom bearings, as well as the temperature at five different locations. If one of these vibrations or one of these temperatures exceed their threshold (safe\makered{ty} limit) for a set duration the system is brought to a halt and the coolers/heaters are switched off. Each of the cylinders can be mechanically locked. These locks can be in a `lock' position or a `free' position, both are actively detected using limit switches by Eaton (type: \texttt{LSE-11}). The signals for each pair of these sensors go through \texttt{XOR} gates; the entire system will only operate if the output of these gates are both high. If the `lock' position is detected the system will ensure that the respective motor cannot be operated in order to prevent damage to belts, engines, and pulleys. The bottom plate of the outer cylinder has toothed edges \makered{which are detected by two proximity sensors.} These proximity sensors by Omron (type: \texttt{M8 shielded}) are positioned and the size of the teeth is chosen in such a way that \makered{at least one of the sensors always detects} a metal tooth. The signals of both sensors go through an \texttt{OR} gate, and only if this output is high the system is allowed to operate. This protective feature ensures that the outer cylinder is present during operation. The entire system is housed inside 4 metal frames with strong\makered{,} polycarbonate windows. The front panel has two doors to access the apparatus. The software is configured such that the system is allowed to rotate only if the doors are closed, and the system will halt if the doors are forced open during operation. The system can be halted in the case of an emergency by the safety module, or indirectly by the LabVIEW software (either automatically or manually), but in addition the system also has three emergency buttons that are within the vicinity of the operator. \makered{The buttons all cause the power of the engines to be interrupted.} Moreover, the system is also connected to the safety features of the laboratory. In case the apparatus is operated without the outer polycarbonate frames (this might be necessary if more space around cylinders is required for a laser or camera equipment), the door of the lab cannot be opened from the outside while the system is in rotation---minimizing the risk that someone can come in direct contact with the outer cylinder while the system is operational. Running the apparatus without front panel and doors will enable another layer of security: the system will make a beeping sound for a few seconds before the rotation is started. This last feature ensures that everyone is aware that the system is about to rotate and stays away from the rotating parts. The front panel (or a separate pole that is needed for operating the system without front panel) also features three lights and a sound alarm that indicate the current status: ready to operate (green light), operating (orange light), system halted \makered{due to} the safety mechanisms (red light + sound signal), see also the rendering of the entire apparatus in fig.~\ref{fig:rendering}. 
 
\subsection{Other details and features}
\begin{figure*}[htpb!]
	\begin{center}
		\includegraphics{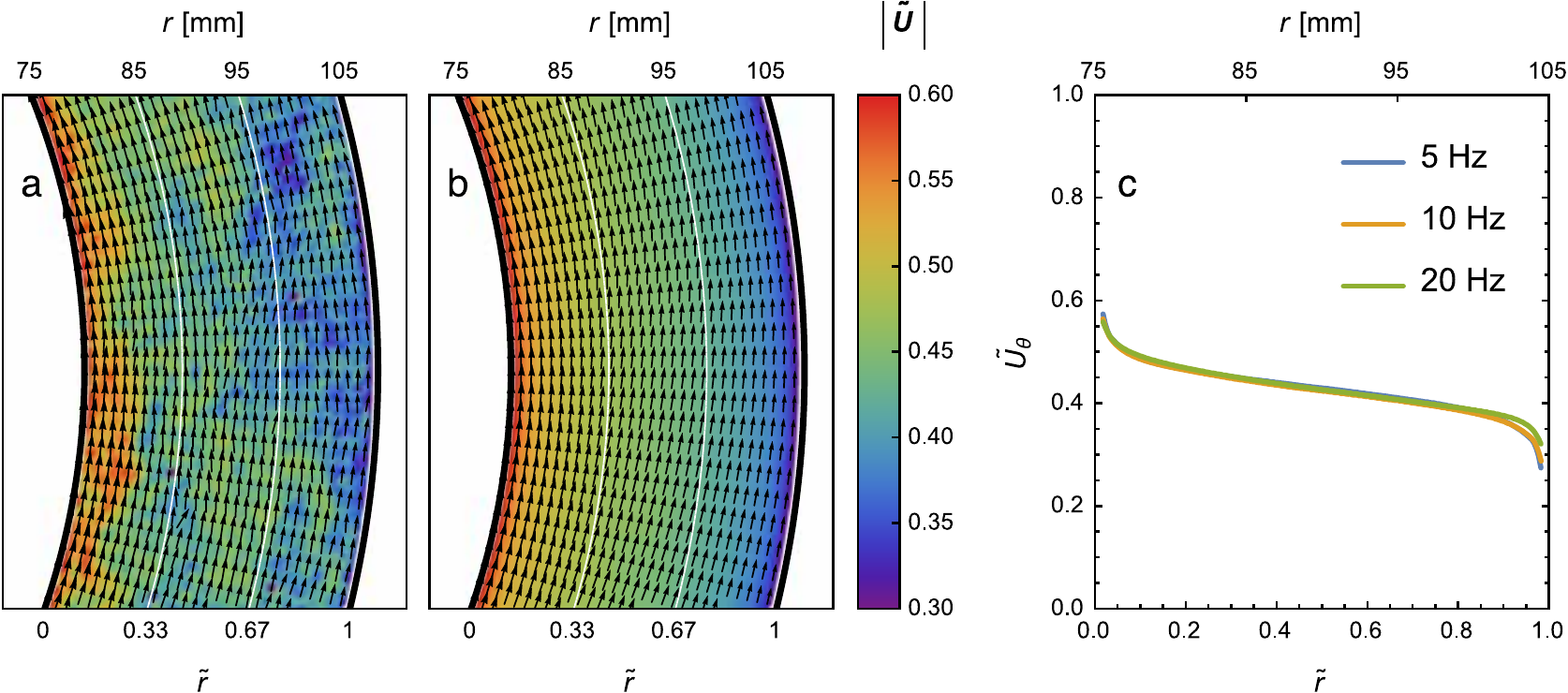}
		\caption{(a) Instantaneous flow field for a rotation of $f_i = \unit{10}{\hertz}$ and stationary outer cylinder. Color indicates the magnitude of the vectors. We define $\tilde{r}$ as the normalized distance between the cylinders: $\tilde r = (r - r_i)/d$, and the normalized velocity as $\left|{\tilde{\mathbf{U}}} \right| = \left|\mathbf{U} \right|/(2\pi f_i r_i)$, and $\left| \hdots \right|$ denotes the Euclidean norm. (b) Averaged flow field for the same rotation velocities. (c) Radially binned, and temporally and azimuthally averaged azimuthal velocity profiles normalized with the driving velocity ($\tilde{U_\theta} = U_\theta/(2\pi f_i r_i)$) for $f_i = \unit{5}{\hertz}$, $f_i = \unit{10}{\hertz}$, and $f_i = \unit{20}{\hertz}$, with Reynolds numbers $\text{Re}_i = 0.7 \times 10^5$, $\text{Re}_i = 1.4 \times 10^5$, and $\text{Re}_i = 2.8 \times 10^5$, respectively. Due to insufficient optical resolution we are unable to measure details of the boundary layer accurately in this measurement, but with modified optics focusing on the boundary layers this would be possible, too.}
		\label{fig:piv}
	\end{center} 
\end{figure*}

The system is filled from the bottom of the apparatus through two filling channels embedded inside the bottom plate of the outer cylinder. The system can overflow through a special channel at the top of the apparatus; the liquid of the measurement gap is connected to 8 channels above the top inner cylinder where the expanding liquid can escape the system, see fig.~\ref{fig:cross} ``Overflow channel'', or see the green channels in fig.~\ref{fig:cooling}. The system can be emptied at the bottom through the same channels as for filling. Apart from a \makered{transparent} outer cylinder and top plate with cooling channels we also have an outer cylinder and top plate that have no channels and no temperature sensors and are \makered{also} completely transparent. \makered{One can attain clearer images using these parts because the number of interfaces is reduced, which might be imperative for \textit{e.g.} high resolution PIV measurements through the top plate or the outer cylinder.} The inner cylinder also embeds a hollow flanged reaction torque transducer (Althen GmbH, type: \texttt{01167-051}) that can sense the torque that is needed to drive the inner cylinder. The inner cylinder is mounted on bearings to the driving axis, and the torque sensor connects the driving axis with the inner cylinder. \makered{The torque sensor senses the torque on the entire copper inner cylinder, having a height of $L=\unit{489}{\milli \meter}$, see also figure \ref{fig:cross}.} The inner cylinder has several seals to prevent water from leaking into it. It remains to be seen whether or not the shear of the flow is high enough such that the friction of all these seals can be considered as negligible (in order to \makered{solely} get \makered{the fluid torque} signal). The frame features an integrated swinging crane (Demag, type: \texttt{DS-Pro 5-500} with steplessly variable controller) with a jib that can be used to assemble and disassemble the system with ease. The integrated crane is telescopic and can be lowered to provide clearance to move it in and out of the lab, see also fig.~\ref{fig:rendering}. A custom-made optical table (dimensions: $\unit{1500}{\milli \meter} \times \unit{900}{\milli \meter}  \times \unit{60}{\milli \meter}$, made by Thorlabs) installed around the \tc~cylinders provides the means to easily mount cameras, optical elements, and other equipment, see fig.~\ref{fig:rendering}. The optical table, the cylinders, the bearings, the engines, and all the supporting structure are mounted inside the main frame on vibration isolators made out of rubber. The vibrations of the \tc~cylinders, inner frame, and the optical table are isolated with respect to the outer frame and the windows. 

\section{Example of flow measurement}

\begin{figure}[htbp!]
	\begin{center}
		\includegraphics[width=0.75\columnwidth]{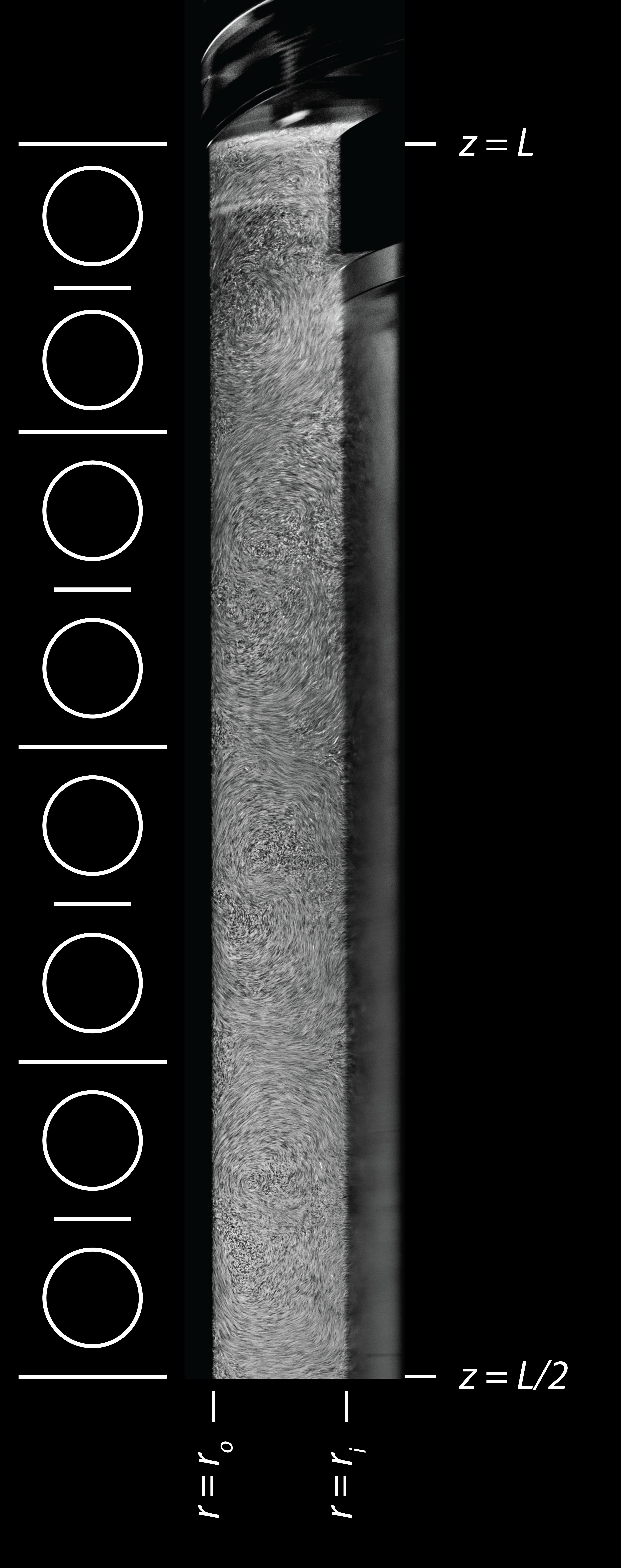}
		\caption{Visualization of the flow in the top half of the radial-axial plane. The flow is illuminated using a laser sheet in aforementioned plane and photographed from the side using a camera. One can clearly see the pattern of the light streaks visualizing the trajectory of tracer particles. Indications on the left show the placements of the vortex pairs.}
		\label{fig:radialaxial}
	\end{center} 
\end{figure}

The BTTC apparatus features a fully transparent top plate allowing for easy flow visualizations in the $\theta$-$r$ plane. This feature is exemplified by performing PIV measurements at three rotational velocities of the inner cylinder ($f_i = \unit{5}{\hertz}$, $f_i = \unit{10}{\hertz}$, and $f_i = \unit{20}{\hertz}$). The fluid is seeded with $1$--$\unit{20}{\micro \meter}$ tracer particles, the fluid (water) is illuminated using a laser sheet around mid height at $z=0.49L$. Based on the particle size and Kolmogorov microscales from Lewis \textit{et al.}\cite{lewis1999} we estimate the particle response time $\tau_p$ and the Kolmogorov time scale $\tau_\eta$. The highest possible Stokes number for our particles (\unit{20}{\micro \meter} particles at the highest rotation speed of $f_i = \unit{20}{\hertz}$, $\rho_p = \unit{1190}{\kilo \gram \per \meter^3}$ as compared to $\rho_f = \unit{1000}{\kilo \gram \per \meter^3}$ \makered{around \unit{20}{\celsius}}) gives $\text{St} = \tau_p/\tau_\eta \approx 5 \times 10^{-4}$. The particles therefore faithfully follow the flow, and can be considered as tracer particles. The laser used is a dual head Nd:YLF laser (Litron, type: \texttt{LDY303PIVHE}). The flow is imaged using an sCMOS camera with a resolution of \unit{5.5}{megapixel} (PCO, Germany) from the top via a front surface mirror mounted at \unit{45}{\degree}. One thousand frames are recorded (with a maximum frame rate of \unit{50}{\hertz}) for each measurement which provides sufficient averaging to become statistically stationary within $1\%$. We perform so-called dual frame PIV measurements such that each \textit{pair} of laser pulses provide two recorded images, which will be processed to a \textit{single} PIV velocity vector field. With the maximum frame rate of \unit{50}{\hertz} of the camera this results in 25 PIV vector fields per second. We perform standard PIV algorithms with adaptive interrogation window size to compute the vector fields. The output of the algorithm are vector fields of $160\times 135$ vectors. The high resolution set of vectors are linearly interpolated and a small set of vectors are shown in a polar grid, see figure~\ref{fig:piv}.

Figure \ref{fig:piv}a shows the instantaneous velocity field around mid-height ($z=0.49L$) for the case of $f_i = \unit{10}{\hertz}$, corresponding to $\text{Re}_i = \makered{\omega_i r_i (r_o -r_i)/\nu = } 1.4 \times 10^5$. The colormap shows the magnitude of the velocity (in the $\theta$-$r$ plane). The flow has been normalized by the rotation speed of the inner cylinder. Because of the high Reynolds numbers we are unable to directly measure the velocity boundary layers---the boundary layers are too sharp and our interrogation window too big to faithfully measure the velocity. The main flow direction is very close to purely azimuthal flow; the radial velocity is relatively low, but is very important in order to get higher-than-laminar transport ($\text{Nu}_\omega>1$ in the terminology introduced in ref.~\cite{eckhardt2007}). Figure \ref{fig:piv}b shows the averaged flow field for the same case as figure \ref{fig:piv}a and displays a flow that is nearly purely azimuthal. In figure \ref{fig:piv}c we plot the azimuthally and time averaged velocity profiles for all the three rotation rates. The velocity has been normalized by the velocity of the inner cylinder: $\tilde{U_\theta} = U_\theta/(\omega_i r_i)$. The boundary layers and the `sharpening' of the boundary layers for increasing driving velocity (Reynolds number) is evident. 

In addition to an outer cylinder which has cooling, also a completely transparent outer cylinder without cooling channels has been manufactured. The completely transparent top plate and outer cylinder without obstructing tension bars allows for direct flow visualization, even for the case of outer cylinder rotation. The flow is visualized in the $r$-$z$ plane by having a laser sheet that is coincident with this plane and that is illuminating tracer particles in said flow. The reflected light from the particles can then be recorded and small illuminated trajectories then visualize the flow direction. The apparatus was set to rotate with a rotation ratio of $a=-\omega_o/\omega_i = 0.4$ and $f_i - f_o = \unit{3.5}{\hertz}$. A high-resolution digital single-lens reflex camera from Nikon (type: \texttt{D800E}, \unit{36}{megapixel}) with a \unit{50}{\milli \meter} lens from Zeiss (type: \texttt{Makro-Planar T* 2/50 ZF.2}) was used to take a photo of the illuminated tracers, see figure \ref{fig:radialaxial}. \makered{P}airs of vortices are marked on the left side. The formation of turbulent Taylor vortices can clearly be seen. The vortices have an aspect-ratio of $\xi = h/(r_o-r_i) = 1.14$, which is in accordance with Ref.~\cite{chouippe2014}. The bottom half of the system is (in a time-averaged sense) a mirrored version of the top half. 

\section{Summary and outlook}
A new \tc~facility with independently rotating cylinders and precise temperature control has been constructed. The performance of the engines and angular encoders are tested and show very good performance for rotation rates ranging from \unit{0.1}{\hertz} to \unit{20}{\hertz}. \makered{The apparatus is capable of reaching Reynolds numbers of up to $\text{Re}_i = \omega_i r_i (r_o - r_i)/\nu =  \pm 2.8 \times 10^5$ and $\text{Re}_o = \omega_i r_i (r_o - r_i)/\nu =  \pm 2 \times 10^5$ for the case of water, and up to $\text{Re}_i = \pm 6.7 \times 10^5$ and $\text{Re}_o = \pm 4.7 \times 10^5$ for the case of FC-3284 liquid. The setup can be heated and cooled from \unit{0}{\celsius} to \unit{60}{\celsius} using water as the coolant and while maintaining a large safety factor. Using a different coolant we can possibly extent this temperature range from \unit{-50}{\celsius} to \unit{80}{\celsius} while maintaining a reasonable safety margin.} The system has a wide variety of safety features to allow for safe operation of the apparatus. PIV measurements are performed and show that it is feasible to quantify the flow structures from the top. Imaging through the outer cylinder using a high-resolution camera allows us to easily measure the turbulent coherent structures in the vertical plane of the apparatus. The setup is able to cool and heat the outer and inner cylinder independently and was found to have a very good temperature stability. This temperature stability will be crucial in order to explore the boiling process in TC geometry, but before embarking on the exploration of phase transitions in Taylor-Couette flows the characterization of single phase flow in the presented geometry is required. Many open questions in the boiling process remain and we plan to study the nucleation process in this well-controlled turbulent environment, and the clustering dynamics of vapor clouds and its effects on nucleation.

\begin{acknowledgments}
We acknowledge the help of Ruben Verschoof, Guenter Ahlers, Martin Bos, and from the staff of TCO (Techno Centrum voor Onderwijs en Onderzoek): Rindert Nauta, Sip Jan Boorsma, and Geert Mentink. This study was financially supported by EuHIT, ERC advanced grant ``Physics of liquid-vapor phase transition'', and a Simon Stevin award by the Technology Foundation STW of The Netherlands.
\end{acknowledgments}


\begin{thebibliography}{10}

\bibitem{tay23}
G.~I. Taylor, Phil. Trans. R. Soc. Lond. A {\bf 223},  289  (1923).

\bibitem{fenstermacher1979}
P. Fenstermacher, H.~L. Swinney and J. Gollub, Journal of Fluid Mechanics {\bf
  94},  103  (1979).

\bibitem{diprima1985}
R. Di~Prima and H. Swinney,  in {\em Hydrodynamic Instabilities and the
  Transition to Turbulence}, Vol.~45 of {\em Topics in Applied Physics}, edited
  by H. Swinney and J. Gollub (Springer Berlin Heidelberg, 1985), pp.\
  139--180.

\bibitem{and86}
C.~D. Andereck, S.~S. Liu and H.~L. Swinney, J. Fluid Mech. {\bf 164},  155
  (1986).

\bibitem{huisman2014a}
S.~G. Huisman, R.~C.~A. van~der Veen, C. Sun and D. Lohse, Nat Commun {\bf 5},
    (2014).

\bibitem{aldredge1996}
R.~C. Aldredge, International Communications in Heat and Mass Transfer {\bf
  23},  1173   (1996).

\bibitem{aldredge1998}
R. Aldredge, V. Vaezi and P. Ronney, Combustion and Flame {\bf 115},  395
  (1998).

\bibitem{ronney1995}
P.~D. Ronney, B.~D. Haslam and N.~O. Rhys, Phys. Rev. Lett. {\bf 74},  3804
  (1995).

\bibitem{berg2005a}
T.~H. van~den Berg, S. Luther, D.~P. Lathrop and D. Lohse, Phys. Rev. Lett.
  {\bf 94},  044501  (2005).

\bibitem{sug08a}
K. Sugiyama, E. Calzavarini and D. Lohse, Journal of Fluid Mechanics {\bf 608},
   21  (2008).

\bibitem{gils2013a}
D.~P.~M. van Gils, D. Narezo~Guzman, C. Sun and D. Lohse, Journal of Fluid
  Mechanics {\bf 722},  317  (2013).

\bibitem{mckinley2015}
S. Srinivasan, J.~A. Kleingartner, J.~B. Gilbert, R.~E. Cohen, B. Milne,
  Andrew~J.\ and G.~H. McKinley, Phys. Rev. Lett. {\bf 114},  014501  (2015).

\bibitem{chandrasekhar}
S. Chandrasekhar, {\em Hydrodynamic and hydromagnetic stability} (Courier
  Corporation, 1961), Vol.~196.

\bibitem{balbus1991}
S.~A. {Balbus} and J.~F. {Hawley}, \apj {\bf 376},  214  (1991).

\bibitem{rudiger2001}
{G. R\"udiger} and {Y. Zhang}, A\&A {\bf 378},  302  (2001).

\bibitem{ji2001}
H. {Ji}, J. {Goodman} and A. {Kageyama}, Mon. Not. R. Astron. Soc. {\bf 325},
  L1  (2001).

\bibitem{hollerbach2010}
R. Hollerbach, V. Teeluck and G. R\"udiger, Phys. Rev. Lett. {\bf 104},  044502
   (2010).

\bibitem{richardzahn1999}
{D. Richard} and {J. Zahn}, A\&A {\bf 347},  732  (1999).

\bibitem{dub05}
B. Dubrulle, O. Dauchot, F. Daviaud, P.~Y. Longgaretti, D. Richard and J.~P.
  Zahn, Phys. Fluids {\bf 17},  095103  (2005).

\bibitem{schartman2012}
E. Schartman, H. Ji, M.~J. Burin and J. Goodman, A\& A {\bf 543},  A94  (2012).

\bibitem{pao12a}
M.~S. Paoletti, D.~P.~M. van Gils, B. Dubrulle, C. Sun, D. Lohse and D.~P.
  Lathrop, A\&A {\bf 547},  A64  (2012).

\bibitem{ohashi1988}
K.-I. Ohashi, K. Tashiro, F. Kushiya, T. Matsumoto, S. Yoshida, M. Endo, T.
  Horio, K. Ozawa and K. Sakai, ASAIO Journal {\bf 34},  300  (1988).

\bibitem{beaudoin1989}
G. Beaudoin and M.~Y. Jaffrin, Artificial Organs {\bf 13},  43  (1989).

\bibitem{ameer1999}
G.~A. Ameer, S. Raghavan, R. Sasisekharan, W. Harmon, C.~L. Cooney and R.
  Langer, Biotechnology and Bioengineering {\bf 63},  618  (1999).

\bibitem{wereley1999}
S.~T. Wereley and R.~M. Lueptow, Physics of Fluids {\bf 11},  325  (1999).

\bibitem{serre2008}
E. Serre, M.~A. Sprague and R.~M. Lueptow, Physics of Fluids {\bf 20},  034106
  (2008).

\bibitem{jeng2007}
T.-M. Jeng, S.-C. Tzeng and C.-H. Lin, International Journal of Heat and Mass
  Transfer {\bf 50},  381   (2007).

\bibitem{lewis1999}
G.~S. Lewis and H.~L. Swinney, Phys. Rev. E {\bf 59},  5457  (1999).

\bibitem{gils2011a}
D.~P.~M. van Gils, S.~G. Huisman, G.-W. Bruggert, C. Sun and D. Lohse, Phys.
  Rev. Lett. {\bf 106},  024502  (2011).

\bibitem{paoletti2011}
M.~S. Paoletti and D.~P. Lathrop, Phys. Rev. Lett. {\bf 106},  024501  (2011).

\bibitem{huisman2013}
S.~G. Huisman, D. Lohse and C. Sun, Phys. Rev. E {\bf 88},  063001  (2013).

\bibitem{huisman2013a}
S.~G. Huisman, S. Scharnowski, C. Cierpka, C.~J. K\"ahler, D. Lohse and C. Sun,
  Phys. Rev. Lett. {\bf 110},  264501  (2013).

\bibitem{ostilla2014}
R. Ostilla-Mónico, E.~P. van~der Poel, R. Verzicco, S. Grossmann and D. Lohse,
  Journal of Fluid Mechanics {\bf 761},  1  (2014).

\bibitem{ostilla2014a}
R. Ostilla-Mónico, R. Verzicco, S. Grossmann and D. Lohse, Journal of Fluid
  Mechanics {\bf 748},    (2014).

\bibitem{ostilla2014b}
R. Ostilla-Mónico, S.~G. Huisman, T.~J.~G. Jannink, D.~P.~M. Van~Gils, R.
  Verzicco, S. Grossmann, C. Sun and D. Lohse, Journal of Fluid Mechanics {\bf
  747},  1  (2014).

\bibitem{cohen1991}
S. Cohen and M.~D. Moalem, Chemical Engineering Science {\bf 46},  123
  (1991).

\bibitem{ameer1999b}
G.~A. Ameer, W. Harmon, R. Sasisekharan and R. Langer, Biotechnology and
  Bioengineering {\bf 62},  602  (1999).

\bibitem{ameer1999ex}
G.~A. Ameer, G. Barabino, R. Sasisekharan, W. Harmon, C.~L. Cooney and R.
  Langer, Proceedings of the National Academy of Sciences {\bf 96},  2350
  (1999).

\bibitem{lakkaraju2011}
R. Lakkaraju, L.~E. Schmidt, P. Oresta, F. Toschi, R. Verzicco, D. Lohse and A.
  Prosperetti, Phys. Rev. E {\bf 84},  036312  (2011).

\bibitem{lakkaraju2013a}
R. Lakkaraju, R.~J. A.~M. Stevens, P. Oresta, R. Verzicco, D. Lohse and A.
  Prosperetti, Proceedings of the National Academy of Sciences {\bf 110},  9237
   (2013).

\bibitem{limbeek2013a}
M.~A.~J. van Limbeek, H. Lhuissier, A. Prosperetti, C. Sun and D. Lohse,
  Physics of Fluids {\bf 25},  091102  (2013).

\bibitem{zhang2014a}
X. Zhang, H. Lhuissier, C. Sun and D. Lohse, Phys. Rev. Lett. {\bf 112},
  144503  (2014).

\bibitem{dhir1998}
V.~K. Dhir, Annual Review of Fluid Mechanics {\bf 30},  365  (1998).

\bibitem{theofanous2002boiling}
T. Theofanous, J. Tu, A. Dinh and T.-N. Dinh, Experimental thermal and fluid
  science {\bf 26},  775  (2002).

\bibitem{dhir2006}
V.~K. Dhir, Journal of Heat Transfer {\bf 128},  1  (2006).

\bibitem{kim2009}
J. Kim, International Journal of Multiphase Flow {\bf 35},  1067   (2009).

\bibitem{zhong2009}
J.-Q. Zhong, D. Funfschilling and G. Ahlers, Phys. Rev. Lett. {\bf 102},
  124501  (2009).

\bibitem{eckhardt2007}
B. Eckhardt, S. Grossmann and D. Lohse, Journal of Fluid Mechanics {\bf 581},
  221  (2007).

\bibitem{gil11b}
D.~P.~M. van Gils, G.~W. Bruggert, D.~P. Lathrop, C. Sun and D. Lohse, Rev.
  Sci. Instrum. {\bf 82},  025105  (2011).

\bibitem{lohse2016arfm}
S. Grossmann, D. Lohse and C. Sun, Annual Review of Fluid Mechanics {\bf 48},
   (2016).

\bibitem{fardin2014}
M.~A. Fardin, C. Perge and N. Taberlet, Soft Matter {\bf 10},  3523  (2014).

\bibitem{lathrop1992}
D.~P. Lathrop, J. Fineberg and H.~L. Swinney, Phys. Rev. A {\bf 46},  6390
  (1992).

\bibitem{schartman2009}
E. Schartman, H. Ji and M.~J. Burin, Review of Scientific Instruments {\bf 80},
   024501  (2009).

\bibitem{merbold2013}
S. Merbold, H. Brauckmann and C. Egbers, Physical Review E {\bf 87},  023014
  (2013).

\bibitem{kerstin2013}
K. Avila and B. Hof, Review of Scientific Instruments {\bf 84},  065106
  (2013).

\bibitem{chouippe2014}
A. Chouippe, E. Climent, D. Legendre and C. Gabillet, Physics of Fluids {\bf
  26},  043304  (2014).

\end{thebibliography}

\end{document}